\documentclass[a4paper,12pt,reqno]{amsart}

\usepackage[centertags]{amsmath}
\usepackage{amsfonts}
\usepackage{amssymb}
\usepackage{amsthm}
\usepackage{newlfont}
\usepackage{stmaryrd}
\usepackage{mathrsfs}

\theoremstyle{plain}

\theoremstyle{definition}

\theoremstyle{remark}

\numberwithin{equation}{section}

  \let\om=\omega

\newcommand{\bbR}{{\mathbb R}}

\newcommand{\opunit}{\text{1}\kern-0.22em\text{l}}

\newcommand{\bsE}{{\boldsymbol E}}

\newcommand{\bsP}{{\boldsymbol P}}

\DeclareMathAlphabet{\mathpzc}{OT1}{pzc}{m}{it}

\newcommand{\rel}{\,|\,}

\newcommand{\id}{\textrm{d}}

\begin{document}

\begin{center}
\noindent{\large \bf Static and Dynamical \\Nonequilibrium
Fluctuations} \\

\vspace{15pt}

{\bf Christian Maes}\footnote{email: {\tt Christian.maes@fys.kuleuven.be}}\\
Instituut voor Theoretische Fysica, K.U.Leuven\\
\vspace{5pt} and\\ \vspace{5pt} {\bf Karel
Neto\v{c}n\'{y}\footnote{email: {\tt netocny@fzu.cz}}}\\
Institute of Physics AS CR, Prague
\end{center}

\vspace{20pt} \footnotesize \noindent {\bf Abstract:} Various notions of fluctuations exist
depending on the way one chooses to measure them. We discuss two extreme cases (continuous
measurement versus long inter-measurement times) and we see their relation with entropy production
and with escape rates.  A simple explanation of why the relative entropy satisfies a
Hamilton-Jacobi equation is added.

\normalsize
\section{Introduction}
All statistical mechanics originates from a fluctuation theory. For equilibrium statistical
mechanics, there is the theory pioneered by Boltzmann, Planck and Einstein where probabilities are
{\it derived} from macroscopic conditions.  So then, if we know say the values of the energy, the
particle number and perhaps a few other macroscopic quantities, we can {\it compute} specific
heats, susceptibilities, etc. and see how they relate to fluctuations in energy, magnetization,
etc. Moreover, thermodynamic potentials appear in the equilibrium fluctuation law to weigh the
deviation from the macroscopic condition.\\
One would hope to achieve the same for nonequilibrium systems.
Things are however more complicated here.  To reach a
nonequilibrium steady state, the system must be sufficiently open
to an external world that permits the flow of energy or particles.
As a consequence, there is no immediate analogue of the
microcanonical ensemble from which all other equilibrium ensembles
are derived. True, we could include that external world in our
system to have again a closed total system but then we must deal
with the problem of relaxation to equilibrium with special and
long-lived
nonequilibrium constraints.\\

In the present note, we deal with aspects of nonequilibrium
fluctuations that, while clearly important, appear less known. One
issue concerns the very notion of fluctuation. We emphasize an
operational interpretation in which the nature of the fluctuation
in fact depends on  how we measure them. Secondly, we point out a
couple of relations between fluctuation rates and thermodynamic or
kinetic quantities.  We explain the relation between relative
entropy and entropy production and we suggest a new way of
computing escape rates in nonequilibrium systems.

\section{Fluctuations}
Fluctuations can be large or small, steady
 or time-dependent; they obviously depend on the
observable and they can e.g.\ measure deviations around the law of large numbers for a macroscopic
quantity.  Yet, in all events, it takes time to measure fluctuations.  But if we speak about
time-averages, we also need to specify how much time we leave after every observation. Especially
so where time-correlations matter, we can expect to see relevant differences between more static
(with large inter-times)
and dynamic (with small inter-times) fluctuations.\\

Imagine we have a time-homogeneous steady state regime and ask what is the probability to see a
particular value for some observable. E.g.\ for given nonequilibrium constraints, you ask how
plausible it is to see a particular (local) density profile. That will be reflected in the
frequency of occurence.  You will probably want to make time-averages over a period $T$ that is
sufficiently long to see a steady value. Yet, for finite $T$ there are always fluctuations. The
point is now that these depend on the way of recording. To be more precise, think of a process
$(X_t, t\geq 0)$, where $X_t$ is the state of your system at time $t$ (position, local density,
magnetization, local kinetic energy,...).  We fix $\tau > 0 $; by a time-average we mean the
average of $X_\tau, X_{2\tau}, X_{3\tau},\ldots, X_T$. Over a long time $T$, we see the value $x$
with a fraction
\begin{equation}\label{ta} p_T^{(\tau)}(x) =
\frac{1}{\lceil\frac{T}{\tau}\rceil}
\sum_{k=1}^{\lceil\frac{T}{\tau}\rceil} \delta_{X_{k\tau} = x}
\end{equation}
($\delta_{a=x}=1$ when $a=x$ and is zero otherwise.) That is the relative frequency to observe $x$
when measuring with time-intervals of length $\tau > 0$, for a duration $T$. There will appear two
extremes: either
$\tau\downarrow 0$ is very small, or $\tau \approx T$.  The first type
of fluctuations will be called \emph{dynamical} and the second case corresponds
to \emph{static} fluctuations. \\

To be simple and to be specific, let us consider
$(X_t)_{t \geq 0},$  a stationary ergodic Markov process, with invariant measure
$\rho$. Have in mind steady state descriptions of mesoscopic
systems via Langevin dynamics, or stochastic lattice gases etc.
 By the ergodicity, for $\bsP-$almost all states,  $\lim_T p_T^{(\tau)} =
\rho$, with $\bsP$ the law of the process.  For finite $T$ it could be that $p_T^{(\tau)}$ very
much resembles a particular probability measure $\mu$.  Or, fixing an arbitrary probability law
$\mu$, we can ask how likely it is to observe it as the statistics of
$(X_t)_t$. The theory of Donsker and Varadhan \cite{DV,DZ}
gives an expression of the rate function
\begin{equation}\label{itau}
  I_\tau(\mu) = -\inf_{g > 0} \sum_{x,y} \mu(x) \log\frac{\sum_{y}
  g(y) P_\tau(y,x)}{g(x)}
\end{equation}
where $P_\tau(x,y)$ are transition probabilities and the infimum is over all
$g > 0$, appearing in the (logarithmic sense) limit $T\uparrow +\infty$ for \eqref{ta}:
\begin{equation}\label{ds}
\bsP[p_T^{(\tau)}\simeq \mu] \simeq e^{-T/\tau\;
I_\tau(\mu)},\qquad T\uparrow +\infty
\end{equation}
Obviously, we can find the stationary distribution $\rho$ from
minimizing $I_\tau(\mu)$ over $\mu$.  That gives a variational
characterization of $\rho$.  Here is a special (limiting) case on
which we will say more in Section \ref{dvsect}.  Taking the limit
$\tau \downarrow 0$ corresponds to successive measurements
with outcomes appearing with frequency
 \[
p_T^{(0)}(x) \equiv p_T(x) = \frac 1{T}\,\int_0^T \delta_{X_t=x}\,\id t
\]
i.e.\ a standard time-average.
 Then,
 \begin{equation}\label{ff}
\bsP[p_T\simeq \mu] \simeq e^{-T I(\mu)}
\end{equation}
where $I(\mu)$ is the Donsker-Varadhan functional that depends on the transition rates of the
Markov process in the following sense:
\begin{equation}\label{mini}
  I(\mu) = \lim_{\tau \downarrow 0} \frac{I_\tau(\mu)}{\tau} = \sum_{x,y} \mu(x)\, k(x,y) -
  \inf_{g>0}\sum_{x,y}\frac{g(y)\mu(x)}{g(x)}k(x,y)
\end{equation}
In fact, this is a special case of a more general formula valid far beyond the restriction to
finite state processes, namely,
\begin{equation}\label{dv}
I(\mu) = -\inf_{g>0}\bsE_\mu \Bigl[\frac {L g}{g} \Bigr]
\end{equation}
The expectation is over the process starting in the distribution
$\mu$ and $L g$ is the generator of the process acting on a
function $g$ (over which we vary in \eqref{dv}).\\
The variational principle originating from~\eqref{mini}-\eqref{dv} appears rather `unknown:'
finding the stationary $\rho$ amounts to finding the minimum of
$I(\mu)$.  Of course, even to evaluate
$I(\mu)$ already another variation must be done, that over $g$. We
discuss more of that problem in Section \ref{dvsect}. The next section deals with the opposite
limit $\tau \thickapprox T$ of static fluctuations.

\section{Varying the relative entropy}

 The limit $\tau \uparrow +\infty$ in the time-average \eqref{ta} (with $T\uparrow
+\infty$) corresponds to an infinite time separation between
consecutive measurements. Then,
\begin{equation}\label{con}
\lim_{\tau\uparrow +\infty} I_\tau(\mu) = S(\mu \rel\rho)
\end{equation}
with $S(\mu \rel \rho)$ the relative entropy. For finite state spaces (and supposing that
$\mu(x)=0$ whenever $\rho(x)=0$),
\begin{equation}\label{relen}
S(\mu \rel \rho) = \sum_x \mu(x) \log\frac{\mu(x)}{\rho(x)}
\end{equation}
We recognize the expression as it figures as thermodynamic entropy
or another related thermodynamic potential in Gibbsian descriptions of equilibrium systems.\\

The relative entropy \eqref{relen} also appears often in
treatments of nonequilibrium systems. We present two ways of
looking at that.

\subsection{Relative entropy and entropy production}
It has often been emphasized that the relative entropy is a non-decreasing function for
dissipative dynamics.  Again, in our context of Markov processes on a finite state space,
\begin{equation}\label{trel}
\frac {\id}{\id t}S(\mu_t \rel \rho) \leq 0
\end{equation}
for the time-evolved probability distribution $\mu_t$; $\rho_t =
\rho$ as it is the invariant distribution.  There is however need
for more physical context if ever we want to interpret
\eqref{trel} as
the positivity of the entropy production.\\
One possibility goes as follows.\\

The entropy production is the total change of the entropy in the
world, system plus environment.  Let us consider a transition
$x\rightarrow y$ of the state of the system. We assume we have an
entropy function $S$ for the system, and the change of the entropy in the system is thus
$S(y) - S(x)$.  When the current between states $x$ and $y$ equals $j(x,y)=-j(y,x)$ (possibly depending on
time), then the average instantaneous entropy production rate for the system is
$\frac{1}{2} \sum_{x,y} j(x,y)\big[S(y) - S(x)\big]$. On the other hand, with some transitions $x\rightarrow y$ in the
system, one imagines there corresponds a flow of particles or energy etc.\ to an external
reservoir. These give rise to a change of entropy
$S(x,y)$ in the environment. Hence, the total entropy production
is $\frac{1}{2} \sum_{x,y} j(x,y)\,[S(x,y) + S(y) - S(x)]$. Let us see how to model that with the
limited means of a Markov process with transition rates $k(x,y)$.\\

When the system is described at time $t$ by the probability distribution $\mu_t$, one can take as
its entropy function $S(x) = -\log \mu_t(x)$. The expected rate of change of the system's entropy
is thus
\begin{equation}\label{syse}
\dot S_s=\sum_{x,y}\mu_t(x) k(x,y)\big[-\log \mu_t(y) +
\log\mu_t(x)\big]
\end{equation}
Further, since we are modeling an open system dynamics, the ratio of transition rates
$k(x,y)/k(y,x)$ should correspond to the entropy change in the environment. E.g.\ if the states
$x$ and $y$ carry different energies, $H(x) \neq H(y)$, then transitions
$x\leftrightarrow y$ are only possible when the system exchanges energy with some heat bath. If the latter has
inverse temperature $\beta(x,y)$, then $k(x,y)/k(y,x)= \exp[-\beta(x,y) \{H(y) - H(x)\}]$ and its
logarithm gives the change of entropy in that reservoir.  Therefore, under the distribution
$\mu_t$, the expected rate of change of entropy in the environment is
\begin{equation}\label{rese}
\begin{split}
  \dot S_e &= \sum_{x,y}\mu_t(x) k(x,y)\log\frac{k(x,y)}{k(y,x)}
\\
  &= \sum_{x,y}\mu_t(x) k(x,y)\log\frac{k(x,y)\mu_t(x)}{k(y,x)\mu_t(y)} - \dot S_s
\end{split}
\end{equation}
Upon adding \eqref{syse} with \eqref{rese}, we obtain the total
mean entropy production rate in the distribution $\mu_t$
\begin{equation}\label{tote} \dot S = \dot
S_s + \dot S_e = \sum_{x,y}\mu_t(x)
k(x,y)\log\frac{k(x,y)\mu_t(x)}{k(y,x)\mu_t(y)}
\end{equation}
which is always non-negative.\\
Where do we see the relative entropy \eqref{relen} in these expressions?  When the stationary
Markov dynamics satisfies the condition of (global) detailed balance, then
\[
\frac{k(x,y)}{k(y,x)} = \frac{\rho(y)}{\rho(x)}
\]
and following \eqref{tote}, the rate of total entropy production
then equals
\begin{equation}
\begin{split}
  \dot S &=\sum_{x,y}\big[\mu_t(x) k(x,y)- \mu_t(y) k(y,x)\big]
  \log\frac{\mu_t(x)}{\rho(x)}
\\
  &= -\sum_{x,y}\frac{\id\mu_t(x)}{\id t} \log \frac{\mu_t(x)}{\rho(x)}
\\
  &= -\frac{\id}{\id t} S(\mu_t \rel \rho)
\end{split}
\end{equation}
where in the second equality we have used the Master-equation for the Markov process.  Only when
there is detailed balance, we can interpret~\eqref{relen} as (the negative of)~the `entropy of the
world', the time derivative of which coincides with the (total) entropy production rate. One way
of dealing with that last remark is to consider only closed systems that are already incorporating
various reservoirs and in which the (total) dynamics is detailed balance with respect to some
overall equilibrium distribution.

\subsection{Relative entropy and the Hamilton-Jacobi equation}
Let us assume that the Markov process has states in an interval of the form
$[A/\varepsilon, B/\varepsilon]\subset \bbR$ and rates
\[
k(X,Y)= \frac 1{\varepsilon}\, w(\varepsilon X, Y-X)
\]
We have written capital letters to indicate that states $X,Y$ represent values of a
\emph{macroscopic} observable(s), which is extensive in $\varepsilon^{-1}$ proportional to the
(great) number of particles.
A standard example is (a continuous time version of) the Ehrenfest
urn model. There a rescaled realization is $x \equiv \varepsilon X
\in [0,1]$, possible transitions $r \equiv Y-X =\pm 1$,
and the rescaled rates $w(x,1)= 1-x$, $w(x,-1)=x$.\\

We consider the probability $p(x,t)$ to have a value $\varepsilon
X =x$ at time $t$.  It satisfies the Master-equation
\begin{equation}\label{me}
\varepsilon \frac{\partial}{\partial t} p(x,t) = -\int\, \id r\; w(x,r)\, p(x,t) + \int\, \id r\;
w(x-\varepsilon r,r) \,p(x- \varepsilon r,t)
\end{equation}
We will be interested in the stationary distribution $p(x)$
verifying \eqref{me} with zero left-hand side. Let us assume that
\begin{equation}\label{inse}
p(x) \propto e^{-S(x)/\varepsilon}
\end{equation}
for some function $S$, up to leading order as $\varepsilon \downarrow 0$. The entropy function $S$
describes \emph{static} stationary fluctuations similarly as the relative entropy~\eqref{relen}
does. To see that, observe that when starting the process at time $-\tau$ from state
$a$ with $S(a)=0$, then
\begin{equation}\label{ss}
\bsP[x(-\tau)=a, x(0)=b] \propto e^{-S(b)/\varepsilon}, \quad \tau\uparrow +\infty
\end{equation}
similar to computing \eqref{ds} for $T=\tau$.\\
There is an algorithm to directly compute $S$, obtained by rewriting~\eqref{me} as
\begin{equation}\label{me1}
  \varepsilon \frac{\partial}{\partial t} p(x,t) = -\int\, \id r\, \Bigl[
  1 - \exp\Bigl(-\varepsilon r
  \frac{\partial}{\partial x} \Bigr) \Bigr] \, w(x,r)\, p(x,t)
\end{equation}
setting the left-hand side zero, and inserting formula~\eqref{inse} into \eqref{me1} with
$\varepsilon \downarrow 0$:
\begin{equation}\label{wr}
\int \id r \,w(x,r)\,\Bigl[ 1 -
\exp\Bigl(r\frac{\partial}{\partial x}S(x)\Bigr)\Bigr] = 0
\end{equation}
We thus get a differential equation for the relative entropy!\\
As an example, again with $x\in [0,1]$, $r=\pm 1$, and
$w(x,1)=1-x$, $w(x,-1)=x$, we see the solution $S(x) = x\,\log 2x
+ (1-x)\,\log 2(1-x)$, the relative entropy for a density $x$ with
respect to density $1/2$ in coin tossing.\\

The above scheme (with a somewhat different emphasis) was first
mentioned in a paper of Kubo, Matsuo and Kitahara, \cite{KMK}.  We
show how equation \eqref{wr} can be
seen as a (stationary) Hamilton-Jacobi equation.\\
Formally, if we write
\begin{equation}\label{ha}
  H(x,p) = \int \id r\,w(x,r)\;\bigl[ 1 - \exp(r p)\bigr]
\end{equation}
then~\eqref{wr} takes the form
\begin{equation}\label{hj}
H\Bigl(x,\frac{\partial}{\partial x}S\Bigr)=0
\end{equation}
We have to explain yet that the expression~\eqref{ha} is indeed the Hamiltonian corresponding to
the path-space action. By that we mean that the probability to see the state $x$ at time $t$ when
at time $t_0$ the state was $x_0$ is given by the path integral
\begin{equation}\label{path}
  \bsP(x,t \rel x_0,t_0) = \int_{x_0 \rightarrow x}
  \id\omega\,\exp\Bigl[-\frac 1{\varepsilon} \int_{t_0}^t \id
  s\,L(x(s),\dot x(s))\Bigr]
\end{equation}
where we `integrate' over all paths $\omega = (x(s);\, s\in [t_0,t])$ such that
$x(t_0) = x_0$, $x(t) = x$, and the Lagrangian
$L(x,\dot x) = p\, \dot x - H(x,p)$ is the Legendre
transform of $H$.  The reason for all that is the following.\\
In equation~\eqref{ss} we compute the left-hand side by the path
integral~\eqref{path} and assume that, as $\varepsilon \downarrow
0$, the integral over all paths $\om$ can be approximated by the
contribution of a single dominant path (which is the usual
situation):
\begin{equation}\label{res}
\begin{split}
  \bsP[x(-\tau)=a, x(0)=b] &= \int_{a\rightarrow b} \id\omega\,
  \exp\Bigl[-\frac 1{\varepsilon} \int_{-\tau}^0 \id s\,L(x(s),\dot x(s))\Bigr]
\\
  &\simeq \exp\Bigl[-\frac 1{\varepsilon} \inf_{a\rightarrow b} \int_{-\tau}^0 \id s\,L(x(s),\dot x(s))\Bigr]
\end{split}
\end{equation}
Hence,
\begin{equation}
  S(b) = \inf_{a\rightarrow b}\int_{-\infty}^0
  \id s\,L(x(s),\dot x(s))
\end{equation}
which means that $S$ is Hamilton's principal function for the Hamiltonian
$H$ associated to the Lagrangian $L$.\\


The idea of connecting the Hamilton-Jacobi equation with
stationary fluctuations first appeared in \cite{KMK}, and was
reinvented by \cite{jona} in the infinite-dimensional context. As
we have shown above, these strategies amount to deriving a
differential equation such as \eqref{wr} for the relative entropy
(such as $S$ in \eqref{ss}) from the (path-space) Lagrangian (such
as $L$ in \eqref{path}).

\section{Escape rates}\label{dvsect}
Recall the rate function \eqref{ff}--\eqref{mini} for the
fluctuations corresponding to time-averages with continuous
measurements. In general, this is not so simple to compute
explicitly, unless the Markov process is detailed balanced.
Indeed, one checks that in that case of detailed balance with
$k(x,y)\rho(x) = k(y,x)\rho(y)$, the minimizing $g$ in
\eqref{mini} is given by $
g(x)=\sqrt{\mu(x)/\rho(x)}$.\\

We now give an application of that same formula \eqref{ff} but for
a nonequilibrium situation: to compute the escape rate from a
local equilibrium. To be specific, we consider a well-known
nonequilibrium model with just one particle (or, what is the same,
independent random walkers) hopping on a ring with $N$ sites on
which work is done. The jumps are nearest neighbor but they are
biased in the sense that the rate for jumping clockwise is
$k(x,x+1) = p$ while the rate for jumping counter clockwise is
$k(x,x-1) =q$. It corresponds to an asymmetric random walk on a
ring.  The situation is sufficiently simple to understand that
there is a unique stationary distribution $\rho$ corresponding to
the uniform placement of the walker on the ring: $\rho(x) =  1/N$.
The question is again about fluctuations: what is the rate at
which
the particle will escape from a given subset of the ring?\\
 It is
most interesting to consider a connected subset $A=\{1,\ldots,n\}$ of $n<N$ sites.  We are given a
probability distribution $\mu$ localized on
$A$, i.e., $\mu(x)\geq 0$ for $x\in A$ and $\sum_{x\in A} \mu(x)
=1$.  Saying it differently, there is no mass outside $A$. Following  the fluctuation
formula~\eqref{ff} we want to find the rate $I(\mu)$ in the probability that the particle remains
in $A$ with occupation weights according to
$\mu$ for the time $T$.\\

To solve for that escape rate $I(\mu)$ we must find the minimum in
\eqref{mini}.  However since $\mu$ is non-zero only on $A$, the
only question is to minimize (over $g>0$ on $\{1,\ldots,N\}$)
\[
\sum_{x=1}^n \mu(x)\big[ p\, \frac{g(x+1)}{g(x)} +
q\,\frac{g(x-1)}{g(x)}\big]
\]
Obviously the minimizer satisfies $g(x)=0, x\notin A$ since all terms are non-negative anyway. So
we are left with finding the minimal $g$ for a functional $I'(\mu)$ corresponding to the dynamics
on $A$ for which again $k(x,x+1)=p, k(x,x-1)=q$ except when $x=1$, respectively $x=n$, in which
case the particle can only hop to the right,
 respectively the left.  That new dynamics on $A$ is
however an equilibrium dynamics (when indeed $n<N$). One checks
the condition of detailed balance for the distribution $\rho_A(x)
\propto (p/q)^x, x\in A$. As a consequence we can take the
minimizer $g$ given as
\[
g(x) = \sqrt{\mu(x)\,\Bigl[\frac{q}{p}\Bigr]^x},\qquad x\in A
\]
and the rate function for $\mu$ becomes, exactly, for $n<N$,
\begin{equation}\label{imu}
\begin{split}
  I(\mu) &= p+q-2\sqrt{pq}\,
  \sum_{x=1}^{n-1}\sqrt{\mu(x)\mu(x+1)}
\\
  &= (\sqrt{p} - \sqrt{q})^2 +
  2\sqrt{pq}\,\Bigl[ 1-\sum_{x=1}^{n-1}\sqrt{\mu(x)\mu(x+1)}\,\Bigr]
\end{split}
\end{equation}
The minimum of that $I(\mu)$ is reached for the uniform
distribution $\mu(x)=1/n$ on $A$: we then have \eqref{imu} $ =
(\sqrt{p} - \sqrt{q})^2 + 2\sqrt{pq}/n$. That is the inverse of
the typical time that the particle feels the drive to escape from
the set $A$.  Fixing a time scale by putting $pq=1$, we see that
that escape time increases for $p=q$.\\

Another application of formula \eqref{ff} concerns the relation
with the so called minimum entropy production principle. Quite
generally, the functional $I(\mu)$ is to first order in the
driving and close to equilibrium an affine function of the entropy
production rate. From that, we understand the minimum entropy
production principle for Markov processes whose rates are a small
perturbation from detailed balance rates, see \cite{KM,MN}.
Details will follow in a separate publication, \cite{mn}.

\section{Final comments}
The previous sections have given a short discussion of different
types of nonequilibrium fluctuations.  They vary from static to
fully dynamical fluctuations.  We have discussed a connection
between the relative entropy as rate function for macroscopic
fluctuations and  a
 Hamilton-Jacobi equation.  We have discussed a
scheme to compute escape rates via the Donsker-Varadhan rate function for the dynamical
fluctuations. Each time, we have seen relations with the entropy production. One
final comment is in order however.\\
Recent years have seen a revival of attempts of constructing nonequilibrium statistical mechanics.
It has however not been clear how far we have been moving beyond close to equilibrium
considerations.  In particular, the obsession with entropy production is probably related to the
good experience we have with the situation around equilibrium.  For descriptions further away from
equilibrium, one will also need to deal with fluctuations of time-symmetric observables (which are
not in the sector governed by the entropy production), see also the discussion in
\cite{poincare,maar}. In fact, we expect that the rate functions that have been considered in the
present paper all carry information about time-symmetric currents when these rate functions are
explored beyond first order around equilibrium.

\vspace{5mm}
\noindent {\bf Acknowledgment}\\ K.~N.~is grateful to the Instituut
voor Theoretische Fysica, K.~U.~Leuven  for kind hospitality. He
also acknowledges support from the project  AVOZ10100520 in the
Academy of Sciences of the Czech Republic.

\bibliographystyle{plain}

\end{document}